\documentclass{article}

\usepackage{amsmath,amssymb,amsthm}
\usepackage{graphicx}
\usepackage{caption}
\usepackage{subcaption}
\usepackage{url}
\usepackage{a4wide}

\theoremstyle{definition}

\begin{document}

\title{Can information be spread as a virus? \\Viral Marketing as epidemiological model\thanks{This is a preprint of a paper whose
final and definite form is in Mathematical Methods in the Applied Sciences.
Please cite this paper as: \emph{Rodrigues, Helena Sofia and Fonseca, Manuel José (2016) . Can information be spread as a virus? Viral Marketing as epidemiological model, Mathematical Methods in the Applied Sciences, 39: 4780--4786. DOI:10.1002/mma.3783}} }

\author{Helena Sofia Rodrigues$^{1,2}$\\
{\tt \small sofiarodrigues@esce.ipvc.pt}
 \and Manuel José Fonseca$^{1,3}$\\
{\tt \small manuelfonseca@esce.ipvc.pt}
}


\date{$^1$ \mbox Business School, Viana do Castelo Polytechnic Institute,\\
Portugal\\[0.3cm]
$^2${\text{Center for Research and Development in Mathematics and Applications (CIDMA)}},
University of Aveiro, Portugal\\[0.3cm]
$^3$ Applied Management Research Unit (UNIAG)\\
APNOR, Portugal\\[0.3cm]
}

\maketitle



\begin{abstract}
In epidemiology, an epidemic is defined as the spread of an infectious
disease to a large number of people in a given population within a short
period of time. In the marketing context, a message
is viral when it is broadly sent and received by the target market through
person-to-person transmission. This specific marketing
communication strategy is commonly referred as viral marketing.
Due to this similarity between an epidemic and the viral marketing process and
because the understanding of the critical factors to this communications strategy
effectiveness remain largely unknown, the mathematical models in epidemiology are
presented in this marketing specific field.
In this paper, an epidemiological model SIR (Susceptible-Infected-Recovered) to
study the effects of a viral marketing strategy is presented. It is made a comparison
between the disease parameters and the marketing application, and Matlab simulations
are performed.
Finally, some conclusions are carried out and
their marketing implications are exposed: interactions across the parameters
suggest some recommendations to marketers, as the profitability
of the investment or the need to improve the targeting criteria of the
communications campaigns.

\smallskip


\smallskip

\end{abstract}

%
%

\section{Introduction}

Viral marketing (VM) is a recent approaching markets and communication
with customers and can potentially reach a large and fast audience
\cite{Lans2010}. VM exploits existing social networks by encouraging
people to share product information
and campaigns with their friends, through email or social networks.
This type of communication has more impact in the customer, because
the information was recommended by friends and peer networks,
instead of standard companies. Marketing campaigns that leverage
viral processes of dissemination are considered widely. They have
certain advantages over traditional mass media campaigns,
especially with the cost effectiveness issues and the ability to
reach specific consumer groups \cite{Bruyn2008}.

When a marketing message goes viral, it is analogous to an epidemic,
since involves a person-to-person transmission, spreading within a population.
Using insights from epidemiology to describe the spread of viruses, a mathematical
model of the VM process is proposed. The main aim of this study is to
develop a set of simulation experiments to explore the influence of several
controlled and external factors that could influence viral campaigns.

The structure of this paper is as follows. In Section~2 a theoretical framework related
to VM is exposed, and it is explained the transmission process of a marketing
campaign that turns viral. Then, an epidemiological model is proposed in Section~3,
that reflects the previous theoretical concepts. An interpretation of the mathematical model
in the marketing context is presented.
The relationship between epidemiology, mathematical modeling and
computational tools allows to build and test theories on the development
of a viral message; thus, the numerical simulations using the model
are done in Section~4, as well as the analysis of the graphics obtained.
Finally, conclusions are presented in Section~5 and some future work is proposed.

\section{Viral Marketing - Theoretical Framework}

It is well known that in recent decades, societies have undergone numerous
 formal and conceptual changes in their economic, social, political,
 communication, among others, standards as a result of continuous
 technological innovation that occurs in the most diverse fields.
 It is in this context that the process of digitization of the
 communications media assumes itself as a significant change engine,
 both in terms of interpersonal relations and marketing communications.

Currently, in this context, a co-existence
between the media classified as online and offline has been witnessing, which results in
a hybridization of languages, strategies and techniques of communication,
translated into the key concepts of interaction and bidirectionality
\cite{Lavigne2002}. Under this relation the production and dissemination
of information ceased to be the monopoly of a specific entity to take
a collaborative dimension, where all individuals are connected to virtual
networks, being able to produce and share content, having the
possibility of its direct manipulation \cite{Castells2004}.

Given the ineffectiveness of traditional promotional communication
strategies - from massified and essentially unidirectional nature -
communicators entities have responded with new approaches, such as
VM, which aims to minimize the resistance of consumers
to the promotional messages: establishing a parallelism between the
biological process transmission of a virus between individuals in a
real context, and the process of transmitting a message from an
internet user to another in digital context \cite{Nail2005}.

VM in its strategic line, is directly related to the
variable promotion of the marketing mix, assuming as an alternative
communication technique, embodied in a dissemination process of
promotional content similar to the logic of a viral epidemic.
The spread of the message on a large scale is made possible by
the collaborative action of individuals in virtual networks
\cite{Barichell2010}. The goal is successfully reached when the
message starts to be spread to a relatively small number of
initially targeted persons and then the information is spread
as an epidemic to a large fraction of potential
consumers \cite{Blaszczyszyn2013}.

Also known as Internet Word-of-Mouth (WOM) marketing \cite{Woerdl2008} or
Word-of-Mouth Advertising \cite{Phelps2004}
VM has been gaining more and more fans, from
professionals to researchers, as an alternative strategy to
traditional communication. In this context a divestment
in the traditional media by major advertisers is been observed, and the transfer of
funds to the online marketing actions is increasing \cite{Hinz2011}.
Among companies that use this communication strategy, stand out examples
such as Procter \& Gamble, Microsoft, BMW and Samsung, who take on the
VM as a consumer-initiated activity that spreads the
marketing message unaltered across the market or segment in a limited
period team, mimicking an epidemic \cite{Gardner2013}.
In this perspective, and according to Dobele, Toleman and
Beverland \cite{Dobele2005}, VM offers three main
advantages to a company: it implies a very little investment;
the act of forwarding electronic messages containing advertising
is voluntary and thus may be viewed more favorably; those who
passes the messages will be more likely to know who have
similar interests (more effective targeting).

In this context, the brands are transferring to consumers the power to
communicate and disseminate their promotional messages along its network
of contacts, with a smaller investment and a much higher velocity compared
to the traditional media, either the Internet or on mobile context.
Woerdl \cite{Woerdl2008} has rank as potential benefits of VM:
inexpensive; reaches audiences within a short period of time; fast and
exponential diffusion; voluntary transmission by sender; more effective
targeting; access to diverse audience through social contacts. Concerning
to the risks associated with VM, the authors highlight:
uncontrollable nature, in particular loss over content and audience
reach, few possibilities to measure success and timing; negative WOM
leading to boycott, ruin, unfavorable attitudes; consumers unwilling
to provide referrals unless there is some return; legal emerging issues
have to be considered; consumers may feel exploited, cheated, used.

As key attributes of a successful viral message, we can highlight
features like entertainment, humor, curiosity, surprise, useful
information and relevant content. The more engaging and compelling
is the message, the more likely to spread successfully through a
network of contacts in continuous growth \cite{Dobele2007}.

In a study by Phelps \cite{Phelps2004} the authors identify 28
reasons for forwarding messages. According to the quantification
of the results we highlight seven, in order of importance:
'Because it's fun'; 'Because I enjoy it'; 'Because it's entertaining';
'To help others'; 'To have a good time'; 'To let others know
I care about their feelings'; 'To thank them'. Regarding the
contrary action - not forward messages - this occurs because
of message contents are not in line with the quality standards
and relevance of Internet users. Apart from these reasons, the
authors insist on a basic strategic procedure based on the
importance of selecting targets who will find information
relevant enough to pass-along.

There are countless possibilities of spread of a viral content.
This can be shared through video sharing sites, personal blogs,
groups or discussion forums, or simply being sent by email, as
web page hyperlink, or as attached content. Regarding the profile
of users that can enhance the spread, there is evidence that internet
users, who are more individualistic and/or more altruistic, trend
forward to more online content than others \cite{Ho2010}.

\section{Epidemiological Model}

In this paper, it is presented a \emph{SIR} model:
\begin{quote}
\begin{tabular}{ll}
$S(t)$ & --- susceptible (individuals who can contract the disease);\\
$I(t)$ & --- infected (individuals who can transmit the disease);\\
$R(t)$ & --- recovered (individuals who have been infected and have recovered).
\end{tabular}
\end{quote}

The analysis of a VM campaign can be explained by a standard epidemic model
\cite{Leskovec2007, sofia2013, sofia2014a, Sohn2013}.

These compartments are mutually-exclusive. In this paper, it is assumed that the total population is constant,
which means that $N=S(t)+I(t)+R(t)$. To describe the model is also necessary to present a set of parameters:
\begin{quote}
\begin{tabular}{ll}
$\delta$ & --- contact rate among individuals per period of time;\\
$\tau $& --- probability of a contact between a susceptible and an infected that results \\
&  in disease transmission;\\
$\beta$ & --- infectivity;\\
$\gamma$ & --- recovery rate.
\end{tabular}
\end{quote}

The number of infections among the susceptible population in a period of time
depends on the constant rate ($\delta$) at a given period and the transmissibility ($\tau$)
of the disease given contact. So, the constant $\beta=\delta \tau$ is called the infectivitiy parameter.

The system of differential equations that translates the dynamics of VM is composed by:
\begin{equation}
\label{ode}
\begin{tabular}{l}
$
\left\{
\begin{array}{l}
\displaystyle\frac{dS(t)}{dt}
= -\beta \frac{S(t) I(t)}{N}\\
\\
\displaystyle\frac{dI(t)}{dt} = \beta \frac{S(t) I(t)}{N}-\gamma I(t)\\
\\
\displaystyle\frac{dR_h(t)}{dt} = \gamma I(t)
\end{array}
\right. $\\
\end{tabular}
\end{equation}
subject to initial conditions
\begin{center}
\begin{tabular}{l}
\label{initial_conditions}
$S(0)=S_{0}, \quad  I(0)=I_{0}, \quad R(0)=R_{0}.$
\end{tabular}
\end{center}

\begin{figure}
\centering
\includegraphics[scale=0.5]{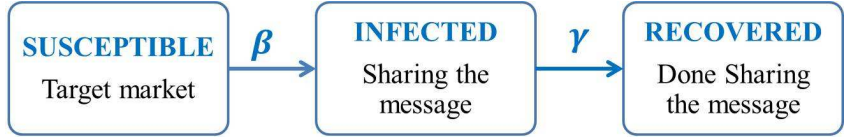}
{\caption{\label{epidemiological_model}  SIR epidemiological model for viral marketing}}
\end{figure}

The scheme of the model is shown in Figure \ref{epidemiological_model}. An arrow pointing into a compartment
is associated with a positive member of the corresponding differential equation while an arrow
pointing out of the compartment represents a negative member of the equation.

The model provides a simple and intuitive approach, modeling the VM
process through epidemiological point of view, where a
disease is spread by person to person contact.
In the marketing context, a susceptible individual is a potential consumer
who may accept the message or
use an offering from a company; this compartment is called the target audience.
An infected individual passes the message or uses a unique product from
the company and/or recommend it. $\beta$ is the probability of moving
from the target audience to infective. The infected individuals start to spread
the message through social contacts. When an individual stops to share a message,
passes to the recovered compartment \cite{Gardner2013, Sohn2013}.

These parameters can also be explained in the marketing context. Infectivity ($\beta$)
is influenced by transmissibility ($\tau$) - through the marketer's efforts to seed the market
with the message and the
acceptability of the cost in time and effort to pass along the message - and the contact rate ($\delta$),
where the extent and the suitability of the social network are key aspects \cite{Gardner2013, Marutschke2014}.
Here, a target member, must perceive the message, has predisposition to deal with the message and finally,
must be motivated to share it. Factors such as exhausting mailing lists, forgetting,
no more interested by this message or
divergent attentions can leave to an individual stop to sharing the message at a rate of $\gamma$,
becoming recovered from the VM campaign. If the exit (recovery) rate is high,
the infective recovers rather quickly.

Thus, it is important to analyze and to assess the impact of a message to the target audience
and the number of members of the target market who have actively shared the message.

\subsection{Model stability and basic reproduction number}

A model reaches its equilibrium when the right side of the differential equations is equal to zero, \emph{i.e.},
\begin{equation}
\label{equilibrium}
\begin{tabular}{l}
$
\left\{
\begin{array}{l}
\displaystyle -\beta \frac{S(t) I(t)}{N}=0\\
\\
\displaystyle \beta \frac{S(t) I(t)}{N}-\gamma I(t)=0\\
\\
\displaystyle \gamma I(t)=0
\end{array}
\right. $\\
\end{tabular}
\end{equation}.

In this model \eqref{ode} we can consider two specific times of equilibrium: one before disease begins spreading
$E_1=(N, 0 ,0)$ and another after disease has moved through entire population $E_2=(0,0,N)$.

In an epidemiology model, the basic reproduction number of the disease, $\mathcal{R}_0$, is
a key concept. The basic reproduction number is the average
number of secondary infections that occur
when one infective is introduced into a completely
susceptible host population (more details can be found in \cite{Hethcote2000, sofia2012, sofia2014b}).

For this model, we obtain
\begin{equation}
\label{R0}
\mathcal{R}_0=\frac{\beta}{\gamma},
\end{equation}

\noindent where $\beta$ is related to the rate that an infected individual
gives rises to new infections and $\frac{1}{\gamma}$ represents the infectious period.
Thus, if $\mathcal{R}_0 <1$, the infection dies out and there is no epidemic, which means that in the
Marketing context the message does not become viral. On other way, if $\mathcal{R}_0 >1$,
the infection will be established in the
population; in terms of business field, the message is spread and all the initial population
have knowledge of the advertising.

\section{Simulations}

The software used in the simulations was \texttt{Matlab}, with the routine \texttt{ode45}.
This solver is based on an explicit Runge-Kutta (4,5) formula, the Dormand-Prince pair.
That means the numerical solver ode45 combines fourth and fifth order methods, both of
which are similar to the classical fourth order Runge-Kutta method. These vary the step
size, choosing it at each step an attempt to achieve the desired accuracy.

\subsection{Parameter Effects}

In this section we made simulations with system (\ref{ode}),
using the following initial values for differential equations:

\begin{center}
\begin{tabular}{l}
$S(0)=900, \quad  I(0)=100, \quad R(0)=0$.\\
\end{tabular}
\end{center}

The analysis of the parameters variation will be focus on the infected compartment -
to visualize the transmission shape - and also on the number of recovered individuals
 at final time, since this population
captures the cumulative number of sharing individuals. The analysis will be done in
dimensionless time, because some campaigns could came viral in days and other in hours or minutes.

\begin{figure}
\centering
\begin{subfigure}[b]{0.45\textwidth}
\centering
\includegraphics[scale=0.45]{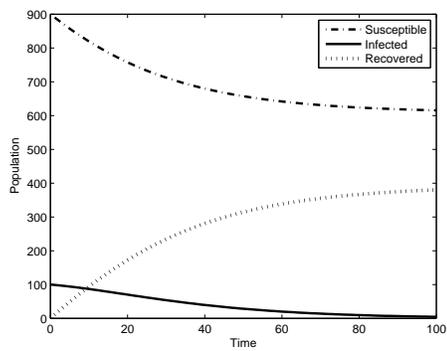}
\caption{$\beta=0.1$}
\end{subfigure}%
\begin{subfigure}[b]{0.45\textwidth}
\centering
\includegraphics[scale=0.45]{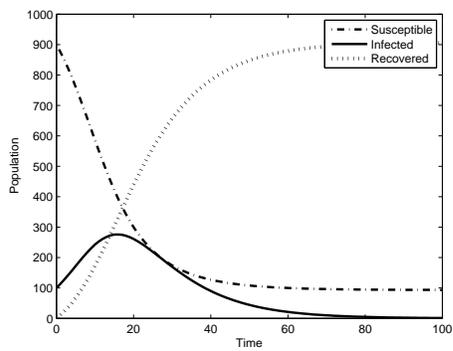}
\caption{ $\beta=0.25$}
\end{subfigure}\\
\begin{subfigure}[b]{0.45\textwidth}
\centering
\includegraphics[scale=0.45]{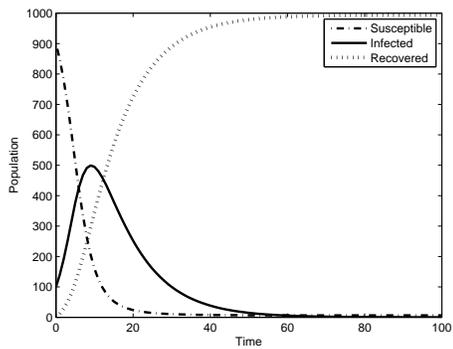}
\caption{$\beta=0.5$}
\end{subfigure}
\begin{subfigure}[b]{0.45\textwidth}
\centering
\includegraphics[scale=0.45]{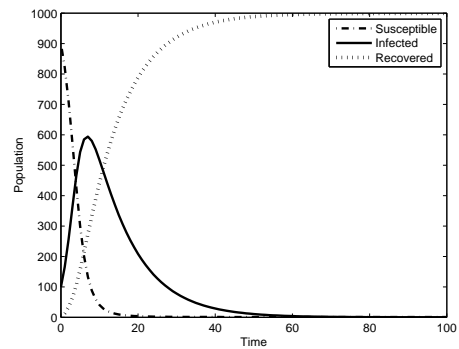}
\caption{$\beta=0.7$}
\end{subfigure}
\caption{SIR model varying infectivity parameter $\beta$ (remaining $\gamma$=0.1)}
\label{beta_analysis}
\end{figure}

Figure \ref{beta_analysis} presents four scenarios for different values of $\beta$,
while the $\gamma$ parameter remains constant ($\gamma=0.1$) and the initial values unchanged.
Increasing the value of $\beta$ implies to reach a peak of transmissibility sooner: in
Figure \ref{beta_analysis}b) (for $\beta=0.25$) is close of time 20 reaching about 300 persons,
while in Figure \ref{beta_analysis}d) the highest value is got close to time 10 and infected about
600 individuals. Observing the recovered curve, is possible to realize that a high value for
infectivity leads to a greater number of target members reached. In the first graphic, the message only
reaches a total of 400 persons of the target audience, approximately. Using $\beta$ at minimum of 0.5,
the viral campaign achieves almost the target audience in less than 60 units of time.

\begin{figure}
\centering
\begin{subfigure}[b]{0.45\textwidth}
\centering
\includegraphics[scale=0.45]{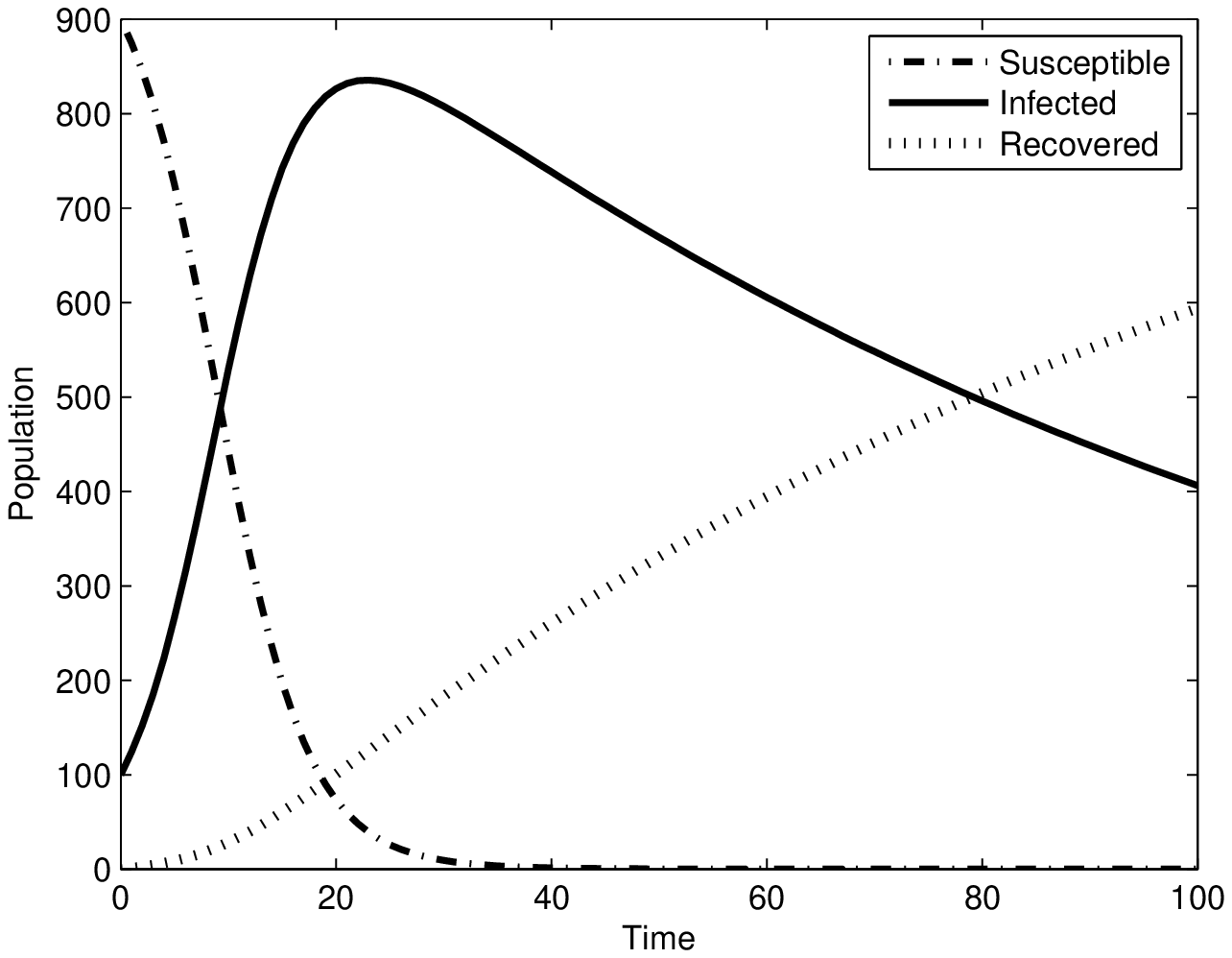}
\caption{$\gamma=0.01$}
\end{subfigure}%
\begin{subfigure}[b]{0.45\textwidth}
\centering
\includegraphics[scale=0.45]{beta025_gamma01}
\caption{ $\gamma=0.1$}
\end{subfigure}\\
\begin{subfigure}[b]{0.45\textwidth}
\centering
\includegraphics[scale=0.45]{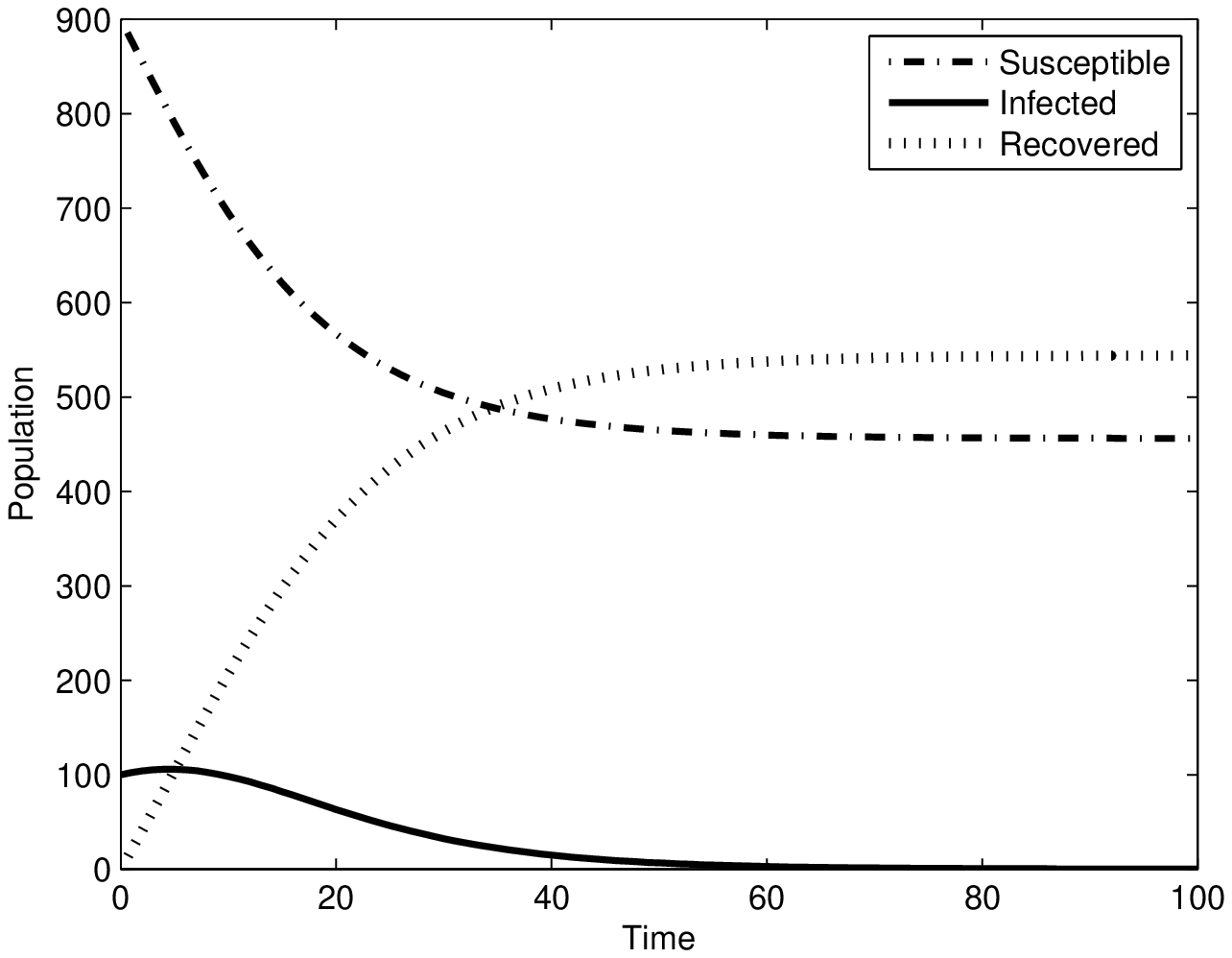}
\caption{$\gamma=0.2$}
\end{subfigure}
\begin{subfigure}[b]{0.45\textwidth}
\centering
\includegraphics[scale=0.45]{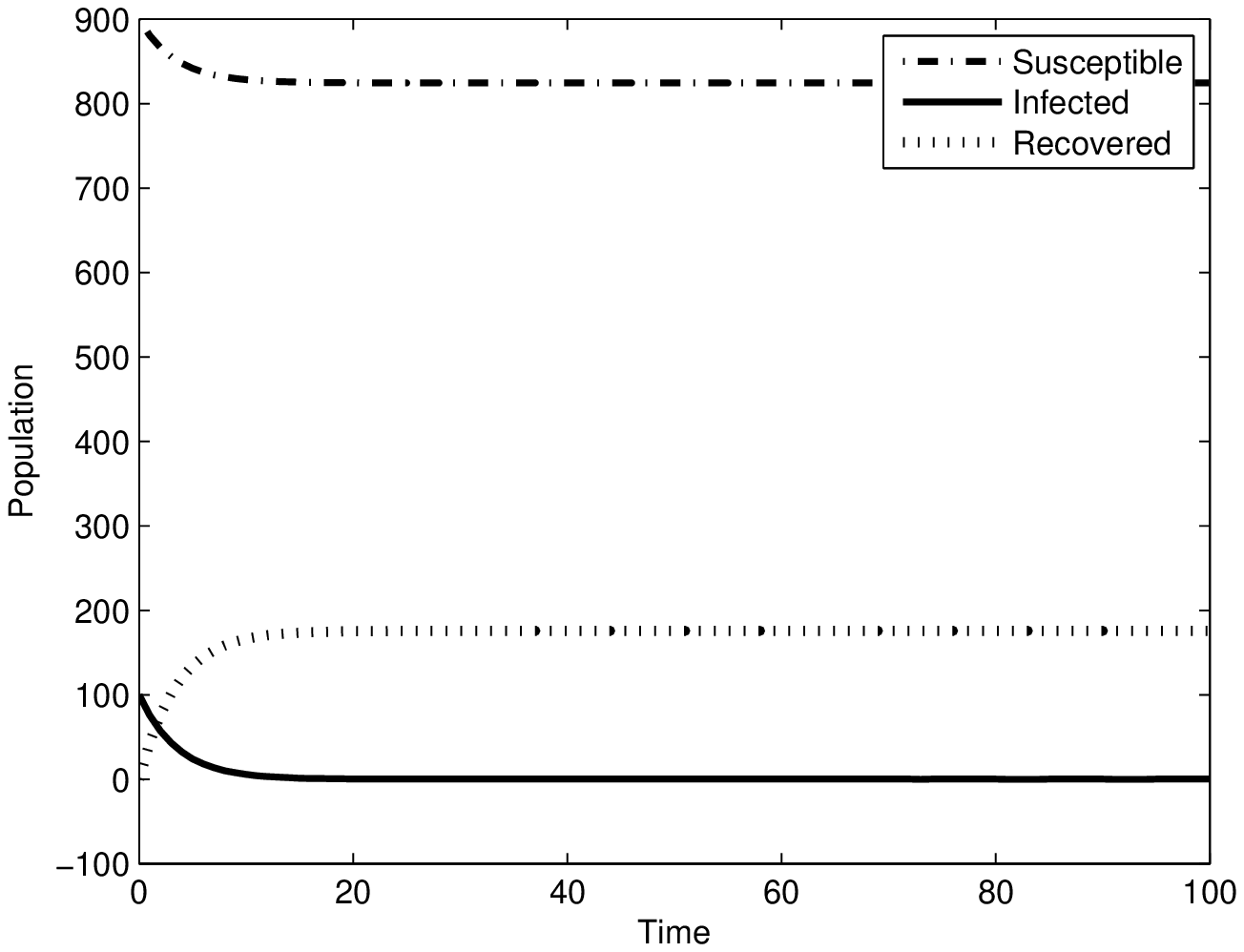}
\caption{$\gamma=0.5$}
\end{subfigure}
\caption{SIR model varying recovery rate parameter $\gamma$ (remaining $\beta$=0.25)}
\label{gamma_analysis}
\end{figure}

Figure \ref{gamma_analysis} is related to the variation of parameter $\gamma$, using the value
0.25 for parameter $\beta$ and remaining the same values for initial conditions. The increasing of $\gamma$
 value leads to a decreasing of sharing message, because an individual tends to
 forget or to be not interested in passing the message, more quickly. In the first graphic ($\gamma=0.01$), all population
is affected for the viral campaign, but in a long period of time; this conclusion can be taken, because in less
than 40 units of time the population ceases to be susceptible. In the last graphic ($\gamma=0.5$), the total of
population reached by the campaign is less than 200 people.

\subsection{Changing Seed Population }

In an epidemic we can not control how many people we have initially infected,
before health authorities start specific treatments or apply measures to control the disease.
However, in a VM message we can simulate and predict the effects of a campaign
depending on the initial behavior of the people that are infected.
One of the challenges of promoting a VM campaign is to know how
many people should
be initially affected, in order to create a huge diffusion of the
product or service.
However, increasing the seeding directly may require more costs for
the company \cite{Lans2010}.

\begin{figure}
\centering
\begin{subfigure}[b]{0.45\textwidth}
\centering
\includegraphics[scale=0.45]{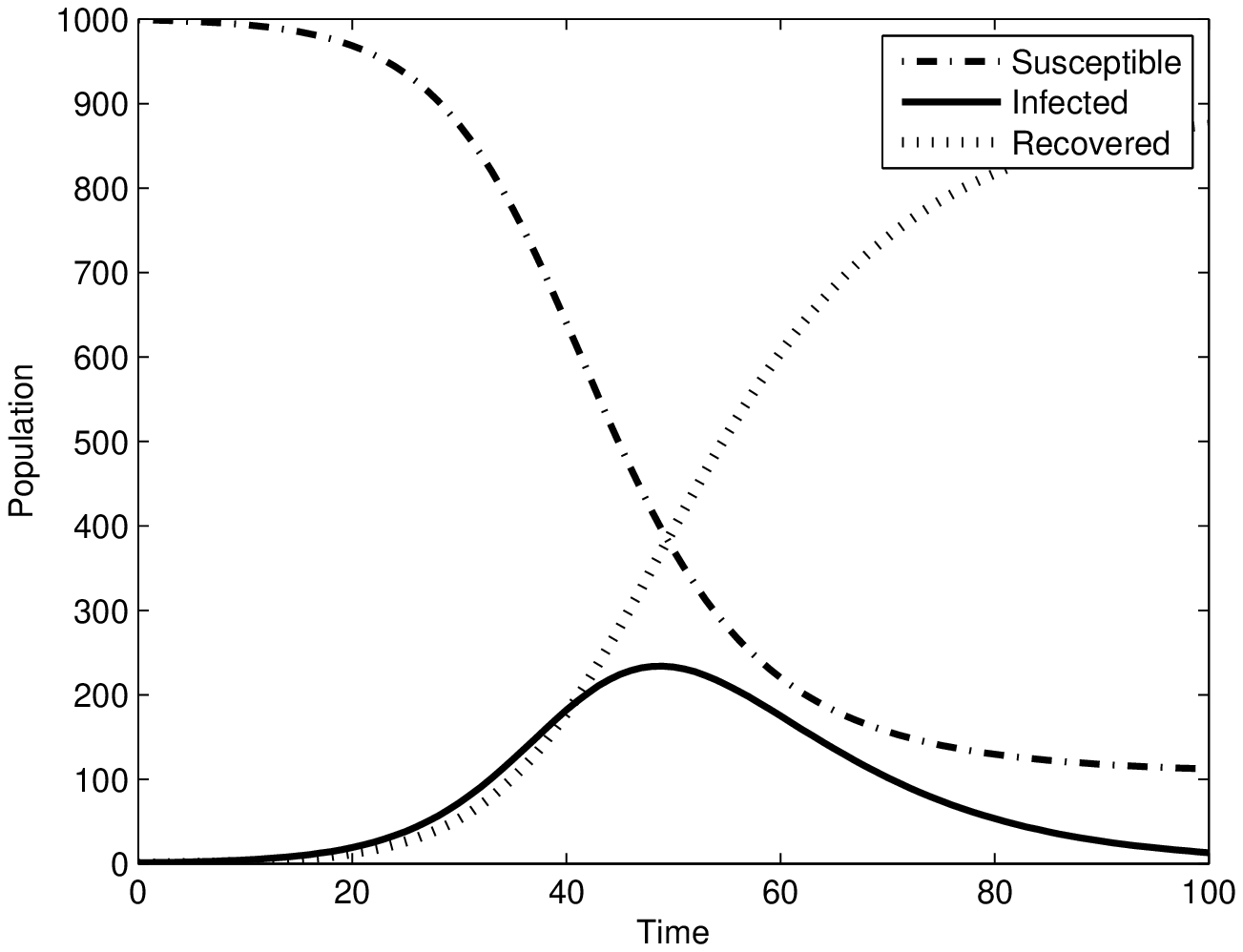}
\caption{$I(0)=1$}
\end{subfigure}%
\begin{subfigure}[b]{0.45\textwidth}
\centering
\includegraphics[scale=0.45]{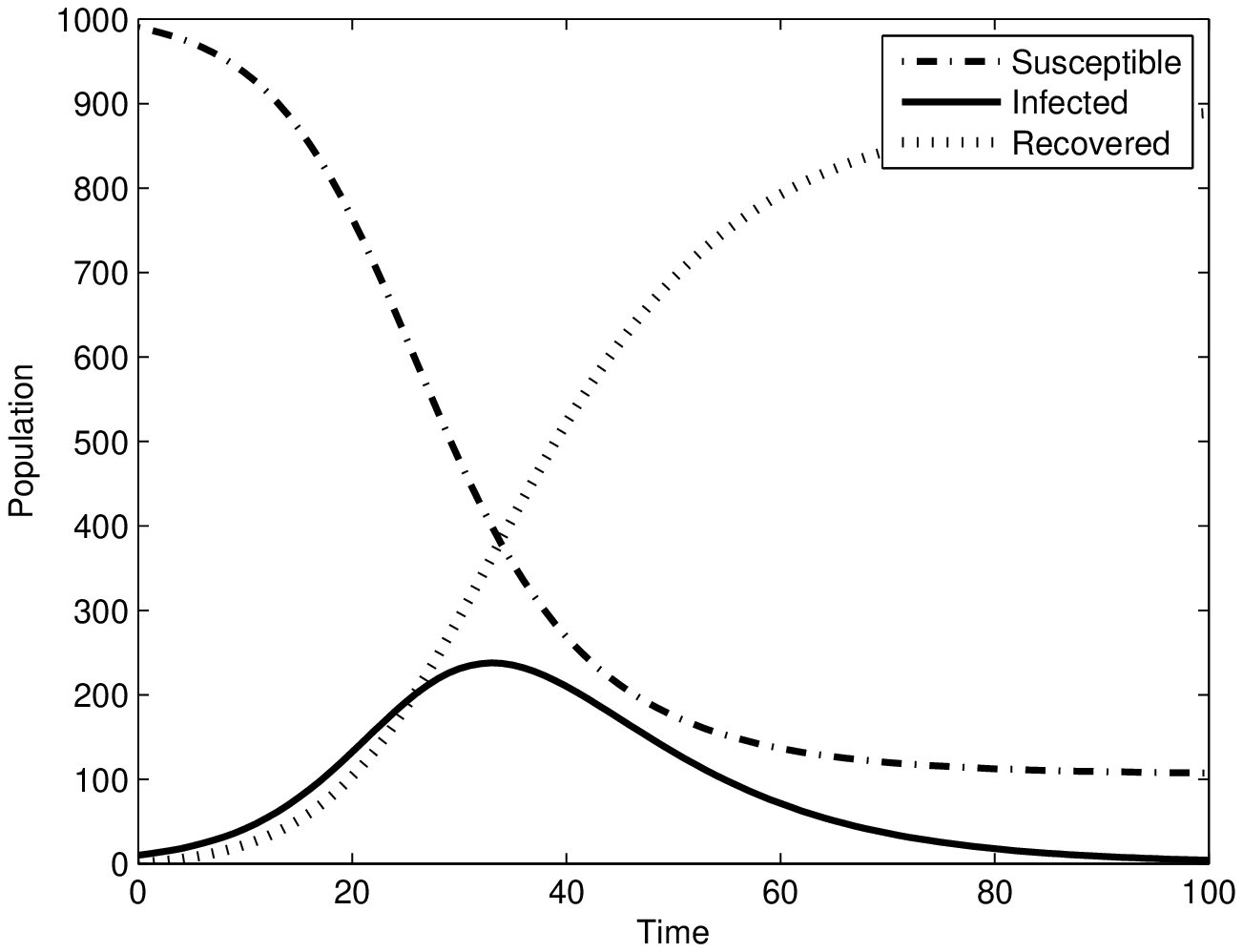}
\caption{ $I(0)=10$}
\end{subfigure}\\
\begin{subfigure}[b]{0.45\textwidth}
\centering
\includegraphics[scale=0.45]{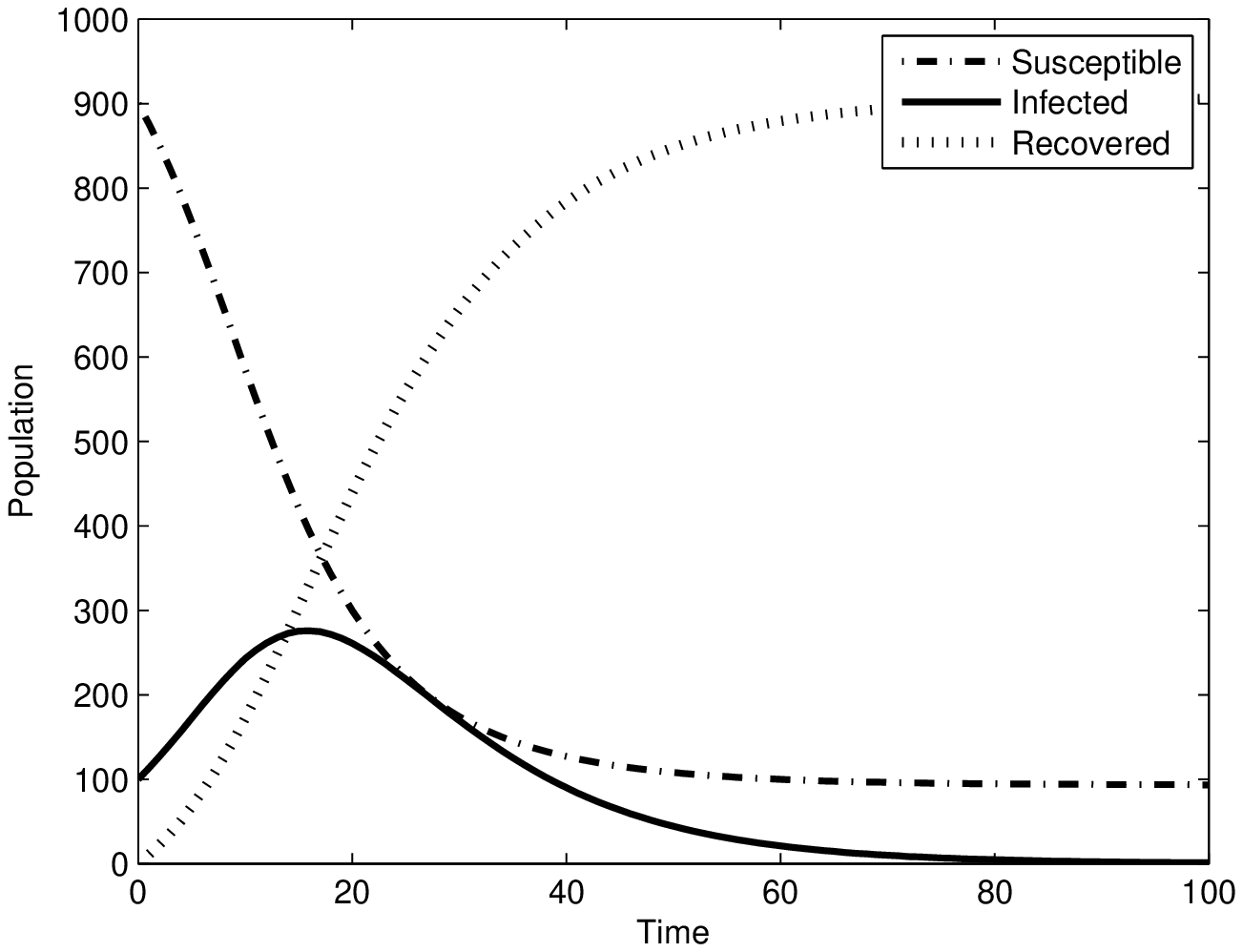}
\caption{$I(0)=100$}
\end{subfigure}
\begin{subfigure}[b]{0.45\textwidth}
\centering
\includegraphics[scale=0.45]{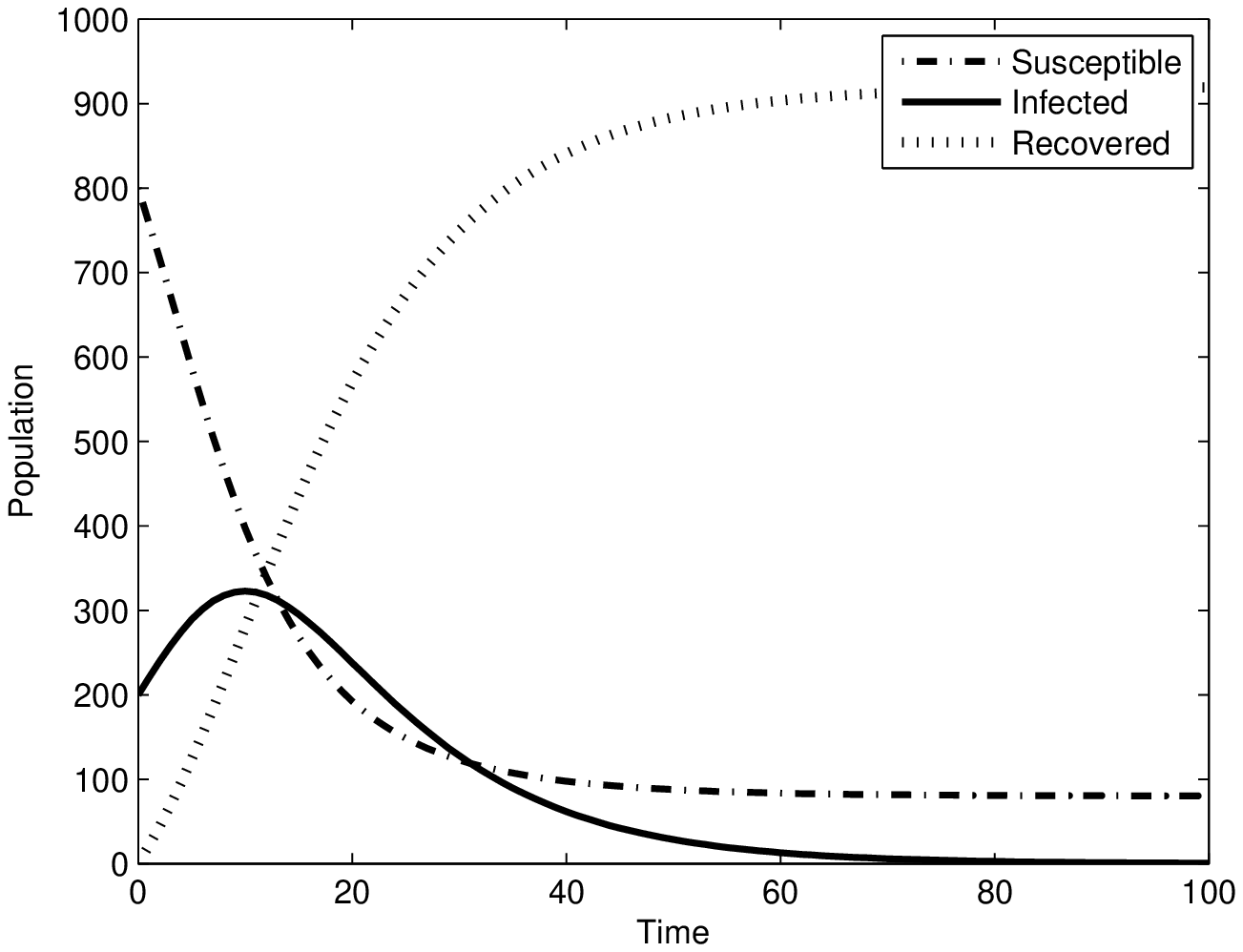}
\caption{$I(0)=200$}
\end{subfigure}
\caption{SIR model varying the initial value of infected individual $I(0)$, remaining $\beta=0.25$, $\gamma=0.1$
and maintaining the population constant, \emph{i.e}, $S(0)=N-I(0)$ and $R(0)=0$}
\label{initial_value_analysis}
\end{figure}

Figure \ref{initial_value_analysis} reflects the variation of
initial value of the target
population that is initial infected. As expected, when the amount
of initially infected people is high,
the viral campaign reaches its peak more rapidly. However, the most
interesting conclusion is that,
the costs of seeding the campaign to 20\% of the target population
produces almost the same effects
(in time and population reached) as invest only 10\% of the target
population to diffuse the campaign.
In most cases, seeding a very large proportion of the population would
likely be very costly in
marketing reality, and the results could not be the expected ones.

\section{Conclusions and directions for future work}

VM as a process, is often portrayed as a random
ground-up phenomenon over which marketers have minor control.
However a more accurate examination makes possible to
identify a number of strategies that can optimize
this marketing communication tool.

This paper presents a model that translate the viral process of a
communications marketing campaign.
The parameters used (infectivity and recovery rate) are very
sensitive. The influence of $\beta$ value
is huge in the message diffusion: as infectivity increases,
the proportion of the
target audience achieved also increases. On other hand, the growth
of $\gamma$ parameter
implies the decreasing of the viral message sharing. In this
way, the marketers
should focus their attention to increase the social network
(increasing $\beta$)
that has specific characteristics for the commercialization of
the product or service,
accurating targeting criteria, in order to
create more interest in each person to pass the message more
often, faster and
for a long period of time (decreasing $\gamma$).

Thus, we believe that in addition to the creative challenge of
produce original and fun virtual messages, it is important that
the company has a deep understanding of their customers, to better
manage issues such as relevance and complexity of the message. If
marketers can access those particular segments and present them
with targeted information directed to their interests, they could
develop into more influential market members.

The investment in seed the message in an initial set of individuals,
must be studied for each case.
For our simulations, to spend more money in the seed population,
over 10\%, it is fruitless, because the
viral campaign produce similar effects, but with a higher financial
burden for the company.
This measure assumes particular relevance both in an alternative
communication strategy,
such as the VM, as in the context of the current economic status,
where all investments
in marketing communications should be optimized.

As future work we intend to applied this model to specific VM campaigns, in order to
fit our parameters to the reality.

\section*{Acknowledgements}

This work was partially supported by The Portuguese Foundation for
Science and Technology (FCT), through CIDMA (Center for Research
\& Development in Mathematics and Applications) within
project UID/MAT/04106/2013. The authors are grateful to two referees
for valuable comments and suggestions.

%
%

\end{document}